\documentclass[10pt]{article}
\usepackage{graphicx}

\begin{document}
\begin{center}
{\Large \bf NUCLEAR PHYSICS WITH ELECTROWEAK PROBES}
\vskip 1.0 cm 
G. CO'
\vskip 0.2 cm 
Dipartimento di Fisica, Universit\`a di Lecce \\
and, \\ 
Istituto Nazionale di Fisica Nucleare, sez. di Lecce\\
Lecce, Italy
\end{center}
\vskip 1.0 cm 
\centerline {ABSTRACT}
{\small 
The last few years activity of the Italian community 
concerning nuclear physics with electroweak probes is reviewed.
Inclusive quasi-elastic electron-scattering, photon end electron
induced one- and two-nucleon emission are considered. 
The scattering of neutrinos off nuclei in the quasi-elastic 
region is also discussed.
}
\vskip 1.5 cm
\section{Introduction}
\label{sec:intro}
In this paper I present the results obtained by the Italian community
in the years 2002-2004 in the field of the theoretical study of lepton
scattering off medium and heavy nuclei
(A$>$4)~\cite{alb02}-\cite{lal04}.  These results are the product
of numerous collaborations with many foreigner colleagues, their
number is about the same of that of the Italian authors.

The range of the problematics covered by the various publications is
wide, and I have organized my presentation as follows.  First, I shall
discuss some general issues concerning the lepton-nucleus interaction.
Then I shall present the results of inclusive electron scattering,
total photon absorption and those obtained by studying one- and
two-nucleon emission processes.  Last, I shall be concerned about the
application to the neutrino scattering of the nuclear models used to
investigate the electron scattering processes.

Since in writing this article I used numerous abbreviations, 
in order to facilitate
the reader, I give their meaning in Table \ref{tab:acr}. 

\vskip 0.5 cm 
\begin{table}[ht]
\begin{center}
\begin{tabular}{|ll|}
\hline
 FG & Fermi gas \\
 FSI & final state interaction \\
 LDA & local density approximation \\
 LIT & Lorentz inverse transform \\
 LRC & long range correlations \\
 MEC & meson exchange currents \\
 MF & mean field \\
 OB & one body \\
 PWBA & plane wave Borm approximation\\
 RFG & relativistic Fermi gas \\
 RPA & random phase approximation \\
 SRC & short range correlations \\
 WS & Woods Saxon\\
\hline
\end{tabular}
\caption {Acronyms used in the article}
\label{tab:acr}
\end{center}
\end{table}

\begin{figure}[ht]
\begin{center}
\includegraphics[scale=0.5] {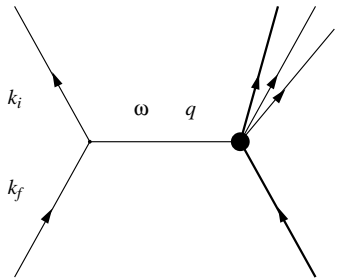}
\caption{One-boson exchange diagram 
  showing the symbols adopted for the
  kinematics variables in the scattering process lepton-nucleus.
  The left vertex represents the lepton, whose four vectors are
  indicated with $k\equiv(\epsilon,{\bf k})$. With $\omega$ and ${\bf
  q}$ I label energy and momentum transfer respectively. The boson
  exchanged are the photon, in the case of the electromagnetic
  interaction, an the $Z^o$ and $W^\pm$ in the case of the weak
  interaction. 
}
\label{fig:feynmann}
\end{center}
\end{figure}

In the study of the lepton scattering off nuclei it is possible to
separate the description of scattering process from that of the
nuclear structure. The first, and quite obvious, reason is that
projectile and scattered lepton are clearly distinguishable from the
hadrons composing the nucleus. In addition, the fact that electroweak
processes are well described already at the first-order perturbation
theory, helps a lot. In effect, all the calculation I have examined
have been done by considering that a single gauge boson is exchanged
between the lepton and the target nucleus (see Fig.
\ref{fig:feynmann}). In addition, also the PWBA has been adopted. With
these approximations, the cross section expressions for both electrons
\cite{bof96} and neutrinos \cite{wal75} scattering processes show a
factorization of the leptonic and hadronic variables.

The leptonic vertex is treated within the relativistic theory, since
the energies involved are much larger than the leptons masses. On
the other hand, the nuclear vertex is usually treated with
non-relativistic quantum many-body theory.

\section{The electron-nucleus interaction}
\label{sec:enucleus}
From now, up to section \ref{sec:nu}, I shall restrict my discussion
to the electromagnetic case. The OB electromagnetic currents are
obtained by summing the currents generated by each nucleon.  Gauge
invariance, i.e. the charge-current conservation law, is not satisfied
if only these currents are considered. This indicates the need of
including other type of currents, produced by the exchange of mesons
between the interacting nucleons and generically called Meson Exchange
Currents (MEC).

\begin{figure}[ht]
\begin{center}
\includegraphics[scale=0.65]{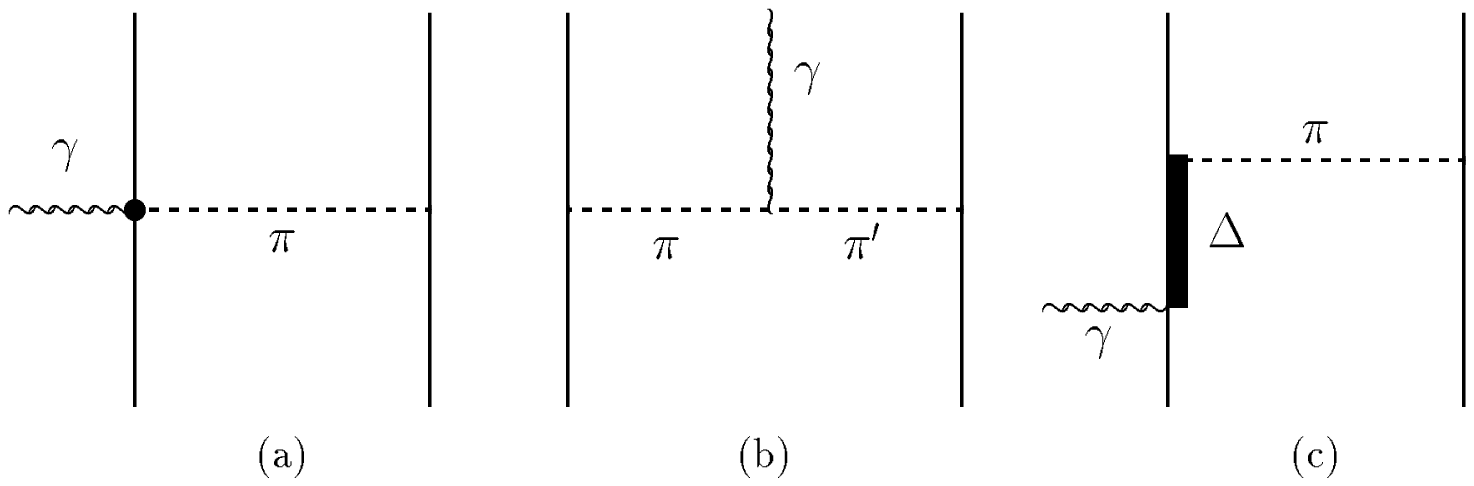}
\caption{Meson Exchange Diagrams considered in the various
  calculations. Contact or seagull (a), pionic or pion in flight (b),
  $\Delta$-current (c).
}
\label{fig:mec}
\end{center}
\end{figure}

Gauge invariance indicates the need of MEC, but it does not define
them in a unique and unambiguous manner. The various methods used to
describe the MEC agree on the fact that the main contributions come
from the three diagrams presented in Fig. \ref{fig:mec}.  The most
relevant terms are the seagull, diagram (a), the and pionic, diagram
(b). They are of the same order of magnitude but they have different
sign. They contribute to the electromagnetic field of the nucleus only
if the exchanged pion is charged. This means that with these two
diagrams only proton-neutron pairs are involved.

At energies far from the peak of the nucleonic 
$\Delta$-resonance the MEC
$\Delta$-current terms, diagram (c), are generally smaller than the
seagull and pionic ones. The $\Delta$-currents contribute to the
electromagnetic field of the nucleus also when the pion exchanged is
chargeless. In this case, the two nucleons involved are of the same
type. This observation is relevant for the two-nucleon emission
processes. 

The validity of the non relativistic reductions used to describe the
electromagnetic field of the nucleus has been studied by investigating
the ideal system of the Fermi gas (FG)~\cite{ama02a}-\cite{bar04}.
The strategy consists in comparing the results obtained for a 
relativistic Fermi Gas (RFG), which is an exactly solvable model, with
those obtained in ordinary non relativistic FG, where various
non-relativistic reductions of the currents have been adopted.

\begin{figure}[ht]
\begin{center}
\includegraphics[scale=0.7]{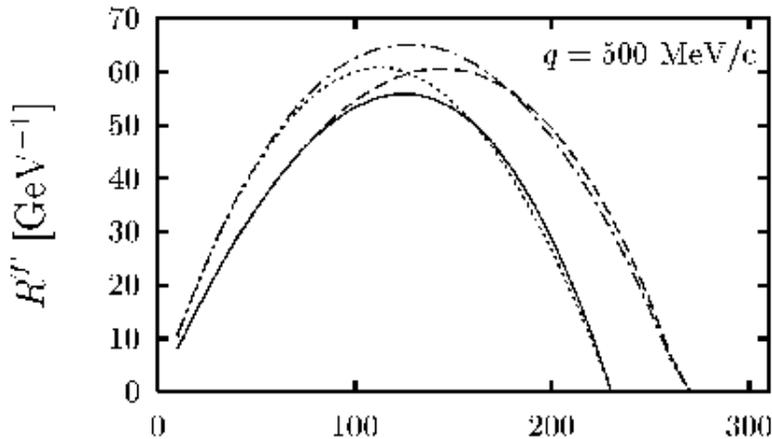}
\caption{Inclusive transverse response. The quantity
  on the x axis is the excitation energy in MeV.
  Full line RFG with OB currents only, dashed
  line FG with OB only, dotted line RFG with OB and MEC, dashed dotted
  line FG with OB and MEC.
}
\label{fig:quique}
\end{center}
\end{figure}

An example of the results of this investigation~\cite{ama02b} is given
in Fig. \ref{fig:quique} where the RFG results obtained with and
without MEC are compared with the analogous results obtained in non
relativistic FG.  This figure shows the transverse response as a
function of the nuclear excitation energy for a given value of the
momentum transfer.  In these results the effect of the relativity is
relatively large.  The height of the peak is reduced by about 
20\%, and also the width of the response is reduced by relativity.
On the other hand, the effect of the MEC does not seem to be sensitive
to relativity. The shift produced by the MEC on the OB responses is
about the same on both RFG and FG results.

\begin{figure}[ht]
\begin{center}
\includegraphics[scale=0.5]{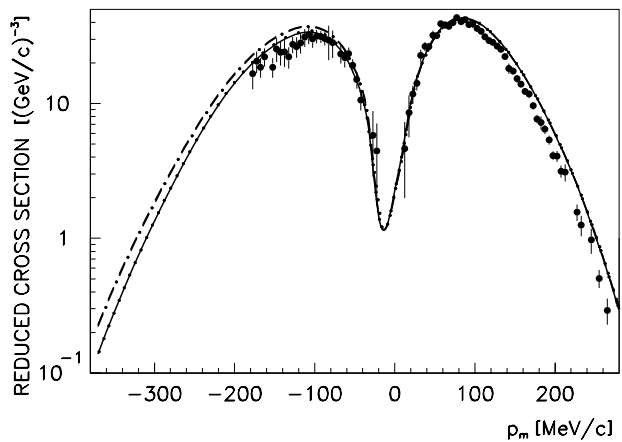}
\caption{Reduced cross section of  $^{16}$O(e,e'p)$^{15}$N calculated
  in the mean field model, as explained in the text, and compared with
  experimental data~\protect\cite{leu94}. The full line has been
  calculated by using all the MEC disgrams shown in
  Fig. \protect\ref{fig:mec}. The result obtained with OB current
  only is shown by the dotted line almost exactly overlapping the full
  line. The dot dashed line shows the result obtained by adding to the
  OB current the seagull diagram. 
}
\label{fig:giustimec}
\end{center}
\end{figure}

The effects shown in Fig. \ref{fig:quique}, are much weaker in finite
nuclei calculations.  In Fig. \ref{fig:giustimec} an example of this
result is seen.  The quantity shown in the figure is the reduced cross
section of the $^{16}$O(e,e'p)$^{15}$N reaction as a function of the
missing momentum~\cite{giu03b}. Also in this case the nuclear wave
functions have been described within a MF model.  A real WS potential
is used to generate the single particle wave functions of the $^{16}$O
ground state.  The parameters have been fixed to reproduce the charge
radius and the single particle energies around the Fermi surface.  The
particle wave function in the nuclear final state, has been obtained
by using a complex optical potential whose parameters have been fixed
to describe the elastic cross sections of the scattering process
between the emitted nucleon and the remaining nucleus with 
A-1 nucleons.

The result of the calculation where all the MEC diagrams of Fig.
\ref{fig:mec} are considered is shown by the full line.  The dotted
line, almost perfectly overlaps the full line, shows the results
obtained with the OB currents only.  The dashed-dotted line has been
obtained by adding to the OB currents only the seagull term. The
results of this calculation show that the contributions of the various
MEC diagrams cancel each other. In this process the effect of the MEC
is so small that they 
cannot be disentangled by a comparison with the data.

\begin{figure}[ht]
\begin{center}
\includegraphics[scale=0.5]{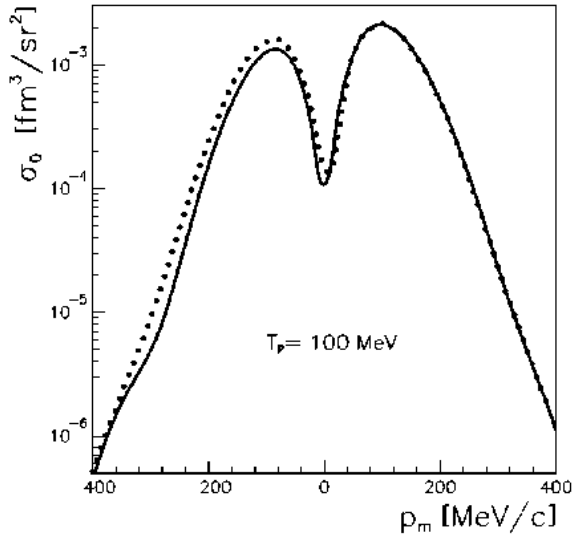}
\caption{Reduced cross section of  $^{16}$O(e,e'p)$^{15}$N. The full
  line shows the result obtained with a relativistic mean field, 
  the dotted line with a non relativistic one.
}
\label{fig:meuccirel}
\end{center}
\end{figure}

The effects of relativity are also strongly reduced in finite
systems~\cite{meu01}.  An example of these results is presented in
Fig.  \ref{fig:meuccirel} where the $^{16}$O(e,e'p)$^{15}$N reduced
cross sections calculated with two different mean field models are
compared.  The full line shows the result obtained with a relativistic
MF, while the dotted line has been obtained with a non relativistic
calculation. Here relativity lowers the height of the maximum by about
10\%. More relevant is the fact that the shapes of the cross sections
are only slightly modified.  Even though Fig.  \ref{fig:quique} and
Fig. \ref{fig:meuccirel} show different quantities, it is evident that
the global effect of relativity is smaller in finite systems than in
FG calculations.  A possible explanation of this can be related to the
procedures used to define the parameters of the nuclear mean fields in
the finite nuclei calculations. In both relativistic and
non-relativistic calculations, these parameters are fixed so as to
reproduce the same quantities. It is plausible that in this fitting
procedure some relativistic effects are effectively included.  This is
not the case of the FG calculations, where the comparison is done
between two ideal systems without any phenomenological parameter to
fix.

\section{Nuclear structure and electron scattering}
\label{sec:ns}
In the previous section, I have discussed some source of uncertainties
in the description of the reaction mechanism between electron and
nucleus. These uncertainties can affect the cross sections on average
by a 10\%, maximum 20\%.  The uncertainties insit in the nuclear
structure produce much larger effects on the cross sections. As a
reminder of this, I would like to briefly recall a set of problems
still open which, however, are not at the moment under the attention
of the community.

\begin{figure}[ht]
\begin{center}
\includegraphics[scale=0.65]{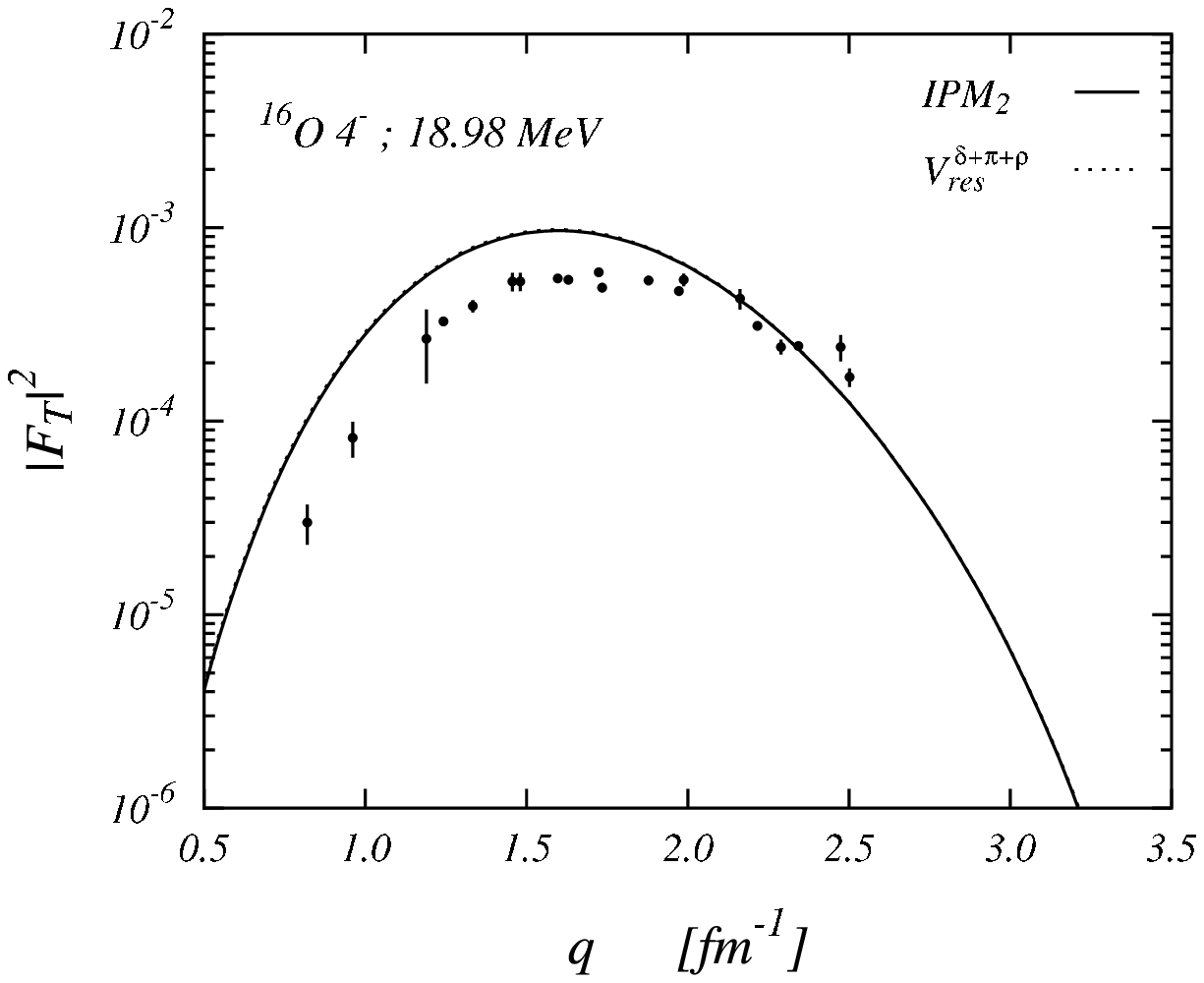}
\caption{Comparison between (e,e') transition amplitudes calculated
  with RPA, full line, and Independent Particle Model, dashed line,
  with experimental data \protect\cite{hyd87} 
}
\label{fig:16o4m}
\end{center}
\end{figure}

(a) 
The high momentum data of the elastic scattering cross sections are
not well reproduced. As a consequence the theoretical charge
distributions are unable to the describe the empirical distributions
in the center of the nucleus. In spite of some attempts~\cite{ang02a}
indicating the physical effects responsible for this discrepancy, to
the best of my knowledge, there is not a single, fully consistent,
calculation able to give a reasonable description of these data.

(b) 
The (e,e') theoretical cross sections in the discrete excitation
usually overestimate the data~\cite{mok00}, as shown, for
example, in Fig. \ref{fig:16o4m}.

\begin{figure}[ht]
\begin{center}
\includegraphics[scale=0.65]{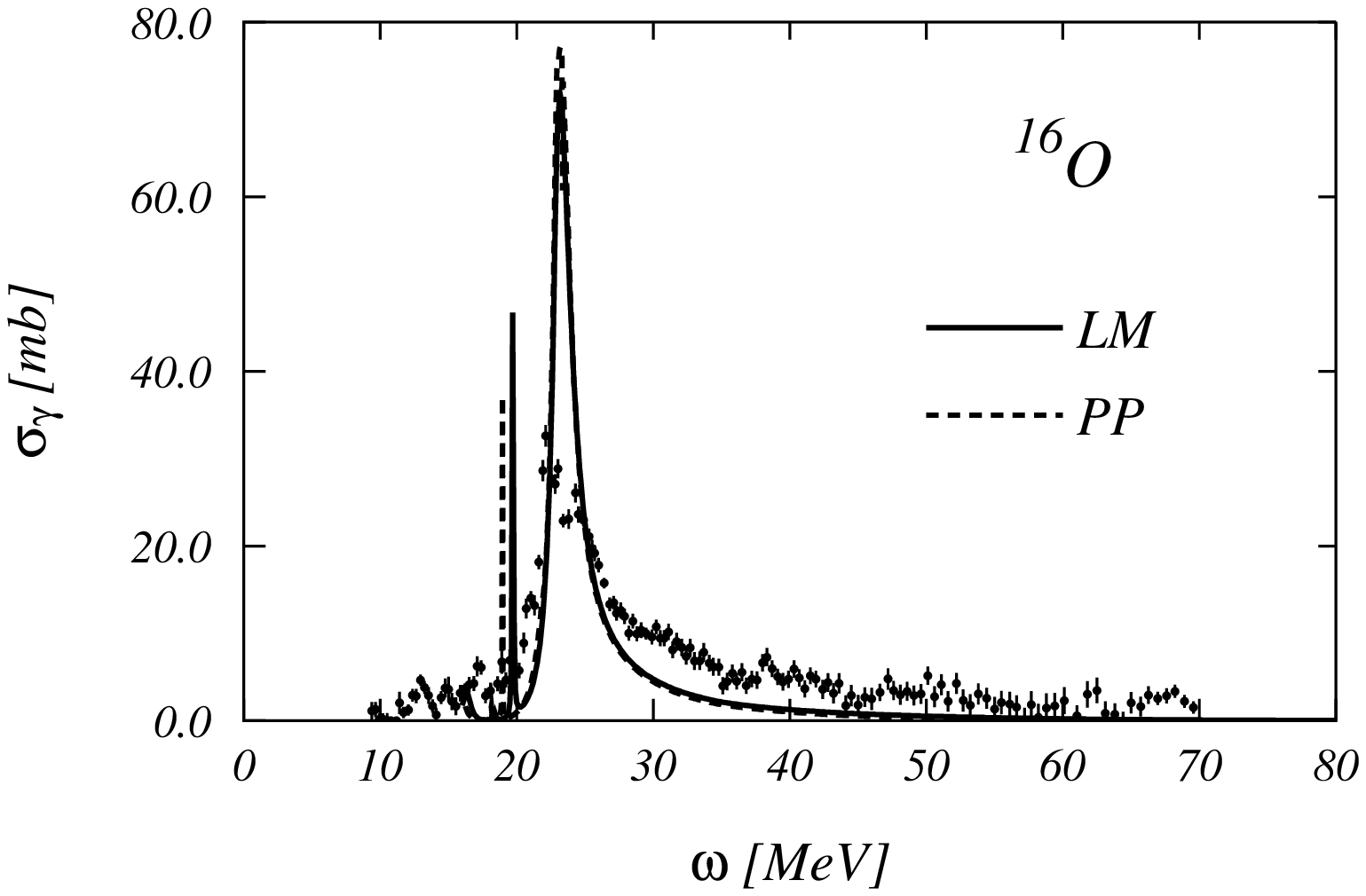}
\caption{Total photon absortpion cross sections calculated within the  
  continuum RPA frwmework with two different residual interactions
  compared with the data~\protect\cite{ahr77}.
}
\label{fig:phot16o}
\end{center}
\end{figure}

(c)
The continuum RPA results for total photo-absorption cross sections in
the giant resonances region, are able to reproduce the energies of the
resonances, but they overestimate the sizes of the cross sections and
underestimate their widths, as shown in Fig. \ref{fig:phot16o}.

The problems I have just mentioned are due to the limitations of the
theoretical models  used to calculate the cross sections
and the other observables. In medium-heavy nuclei the nuclear excited
states are usually described by using the MF model or the RPA. In both
these descriptions only one-particle one-hole excitations, and
eventually their linear combinations, are considered.

These unsatisfactory results are related to the limitations of the
nuclear models adopted and not to the basic assumptions of the theory
used to describe atomic nuclei. This information come from the results
obtained in the few-body systems (A$\leq$4)~\cite{kie04}.  In these
systems the many-body Schr\"odinger equation describing a set of
interacting nucleons, is solved without making approximations, and the
agreement with electromagnetic experimental data is remarkably
superior to what is obtained in heavier systems.

Recently the technique of the Lorentz Inverse Transform (LIT),
previously used in few-body systems, has been applied with great
success to heavier nuclei, up to A=7~\cite{bac02,bac04b,bac04a,orl04}.
A more detailed description of the LIT theory is presented elsewhere
in these proceeding~\cite{qua04}.  Here, I simply want to point out
the fact that this technique accounts for all the possible decay
channels of the nuclear excited state. On the contrary, the MF model,
and also the continuum RPA, consider only the decay in the single
nucleon emission channel.

The MF model is unable to describe low-energy data, but it is quite
successful in the quasi-elastic region dominated by single-particle
dynamics.  In the following subsections I present the results obtained
in this energy region, considering separately the inclusive processes
from those where one, or two, nucleons are emitted and measured in
coincidence with the scattered electron.

\subsection{Inclusive scattering: (e,e')}
\label{sec:ee}
\begin{figure}[ht]
\begin{center}
\includegraphics[scale=1.2]{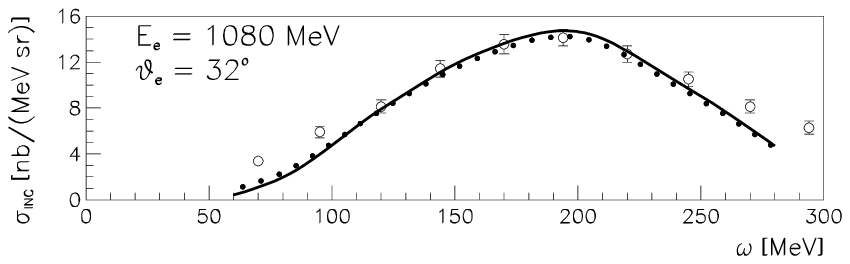}
\caption{Inclusive electron scattering cross section on $^{16}$O
  nucleus calculated in the framework of the relativisti mean-field
  model and compared with the Frascati experimental
  data~\protect\cite{ang96}. The two lines have been obtained by using
  two different approximations.
}
\label{fig:giustiee}
\end{center}
\end{figure}

An example of the agreement between Frascati data~\cite{ang96} on
$^{16}$O and the results obtained with the relativistic MF
model~\cite{meu03} is shown in Fig. \ref{fig:giustiee}.  The same
high-quality agreement is obtained for the other measured kinematics.
The main point of the calculation is the careful treatment of the
Final State Interaction (FSI).  As already discussed in Sect.
\ref{sec:enucleus}, in this MF model, the FSI are taken into account
by using a complex optical potential.  The imaginary part of the
potential removes flux from the elastic channel. In inclusive
experiments the total flux should be conserved.  What is removed from
the elastic channel should go in other decay channels. A more detailed
description of the technique used to conserve the flux is presented
elsewhere in these proceedings~\cite{meu04c}.

The same model~\cite{meu03} has been used to study the separated
longitudinal and transverse responses of $^{40}$Ca and $^{12}$C.  In
$^{40}$Ca the MIT data~\cite{wil97} are well reproduced, while the
comparison with Saclay data~\cite{mez84,mez85} suffers the well
known failures: the longitudinal responses are overestimated, while
the transverse ones are underestimated. The comparison with the
separated responses of $^{12}$C, measured at various values of the
momentum transfer~\cite{bar83}, shows a reasonable agreement with the
longitudinal responses while the transverse ones are always heavily
underestimated.  Old calculations of the $^{12}$C responses done
within a non relativistic framework~\cite{cap91} produce very similar
results. This indicates that the effect of the relativity is
negligible.

The Frascati (e,e')~\cite{ang96} data have been studied by using a
different technique~\cite{ama94}. The basic nuclear model is again the
non-relativistic MF. In this case the responses have been calculated
with a real potential. The results obtained have been folded with
Lorentz functions whose parameters have been fixed to reproduce the
energy behavior of the volume integrals of the optical potential.  In
spite of the technical differences, this approach contains the same
physics as that of the Pavia group, and the results obtained are very
similar.  Also in this case~\cite{co02} the Frascati data~\cite{ang96}
are rather well reproduced. The same kind of agreement is
obtained~\cite{ama94} with the MIT $^{40}$Ca responses~\cite{wil97}.
The comparison with $^{40}$Ca and $^{12}$C Saclay data, shows the same
problems described above~\cite{ama94,ama93}.  The two different
techniques used to treat FSI produce very similar results.

\subsection{One-nucleon emission: (e,e'N) and ($\gamma$,N)}
\label{sec:eep}
The same MF model used to describe the inclusive data has been
utilized to study the single-nucleon emission processes induced by
electromagnetic probes.  The basic ingredients of the
models are the two MF potentials. A real potential that
describes the ground state of the target nucleus, and a complex
optical potential to treat the emitted nucleon wave function.

\begin{figure}[ht]
\begin{center}
\includegraphics[scale=0.7]{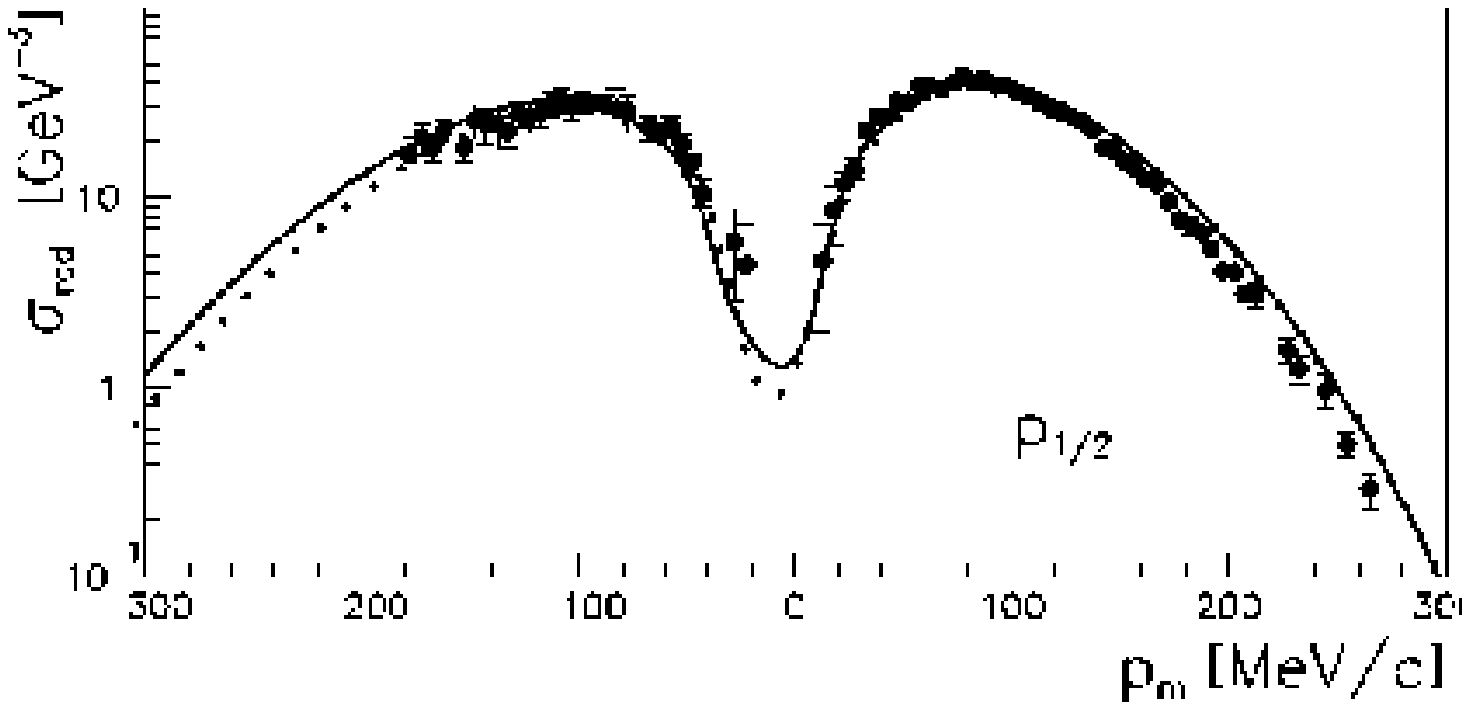}
\caption{Reduced cross section of  $^{16}$O(e,e'p)$^{15}$N compared
  with the experimental data~\protect\cite{leu94}. The full
  line shows the result obtained with a relativistic mean field, while
  the dotted line with a non relativistic mean field. Contrary to
  Fig. \protect\ref{fig:meuccirel} the two results have been multipied
  by two different spectroscopic factors to reproduce the data at best
  (see text).
}
\label{fig:meuccieep}
\end{center}
\end{figure}

The results obtained for the (e,e'p) cross sections in various nuclei
are able to reproduce rather well the behavior of the data after a
quenching rescaling factor is applied~\cite{bof96}.  This rescaling
factor is called spectroscopic factor and it does not depend upon the
kinematics of the experiment.  This is evident since observables
related to the ratio of cross sections are well reproduced without the
use of the spectroscopic factor~\cite{ama03a}. Furthermore, cross
sections measured in very different kinematic conditions are well
reproduced by the same spectroscopic factors~\cite{rad02}.

The spectroscopic factor is a model dependent quantity, as is
deducible, for example, in Fig. \ref{fig:meuccirel}. In this figure the
$^{16}$O(e,e'p)$^{15}$N reduced cross sections calculated~\cite{meu01}
with relativistic (full line) and non relativistic (dotted line) MF
models are compared with the experimental data~\cite{leu94}.  In Fig.
\ref{fig:meuccieep} the full line has been multiplied by 0.7 while the
dotted line by 0.65. This indicates that spectroscopic factors contain
some relativistic effects. These effects are not sufficient to explain
a large part of the spectroscopic factors. It is necessary to go
beyond the MF model, or in other words, to include correlations.  The
investigation of the effects induced by the correlations has been
conducted by using two different approaches.

The basic quantities necessary to calculate the cross sections are the
Fourier transforms of the transition densities induced by the current
operators $ J({\bf r})$:
\begin{equation}
W({\bf q}) = \int d^3 r\, e^{i{\bf q}\cdot {\bf r}}
             <\Psi_f \mid J({\bf r}) \mid \Psi_0 >\,\,\,.
\end{equation}

In the approach adopted by the Pavia group these quantities are 
calculated as:
\begin{equation}
W({\bf q}) = \int d^3 r\, e^{i{\bf q}\cdot {\bf r}} \,\,
             \chi({\bf r})J({\bf r}) \,
   <\Psi^{A-1} ({\bf r})\mid\Psi^A_0 > 
\label{eq:pavia}
\end{equation}
where $\chi({\bf r})$ is the wave function of the emitted, and
detected, nucleon. The important quantity is the overlap function
between the wave function describing the ground state of the target
nucleus and the wave function describing the state of the nucleus
composed by A-1 nucleons.  All the complications related to the
correlations are contained in the overlap function. The formalism
developed by the Pavia group is independent from the methods used to
estimate the overlap function. In the MF model the overlap function is
the single particle wave function of the nucleon below the Fermi
surface.

The approach used in Lecce makes an ansatz on the expression of the
nuclear wave function which is supposed to be the product of a
symmetrized many-body correlation function $F$ and a Slater
determinant.
\begin{equation}
W({\bf q}) = \int d^3 r \, e^{i{\bf q}\cdot {\bf r}}  \,\,
  <\Phi_f \mid F^\dagger J({\bf r}) F \mid \Phi_0 > 
\label{eq:lecce}
\end{equation}
The Slater determinant $\mid \Phi_0 >$ describing the ground state is
composed by all the single particle states below the Fermi surface,
while $\mid \Phi_f >$ contains a hole and a particle states.  The same
correlation function has been used for both ground and excited states.
The many-body correlation function is written as a product of two-body
correlation functions. A cluster expansion is done, and only the terms
containing a linear dependence from the two-body correlation function
are retained~\cite{mok01}. This approach is more tuned to investigate
the so-called short-range correlations (SRC) due to the hard core
repulsion of the nucleon-nucleon potential.

In spite of the differences, the results obtained by the two
approaches are very similar. In general, for the considered processes,
the effects of the SRC are very small.  Certainly the inclusion of the
SRC does not reduce sensitively the values of the spectroscopic
factors.  As an opposite example of the behavior of the correlation,
I like to quote the results obtained in $^{40}$Ca by using overlap
functions produced by the Generator Coordinate Method.  In this case
the spectroscopic factors are even larger than those of the MF model
calculation~\cite{iva02,iva01}.

From the qualitative point of view, the results obtained with the two
methods described above, show that the presence of SRC does not modify
sensitively the shapes of the (e,e'p) cross
sections~\cite{gai02,mok01}.  The differences between cross sections
obtained with and without SRC are within the accuracy of the
experimental data.  An interesting deviation from this general trend
is the case of the $^{32}$S nucleus~\cite{gai02}, which should be
worth further investigation.

\begin{figure}[ht]
\begin{center}
\includegraphics[scale=0.7]{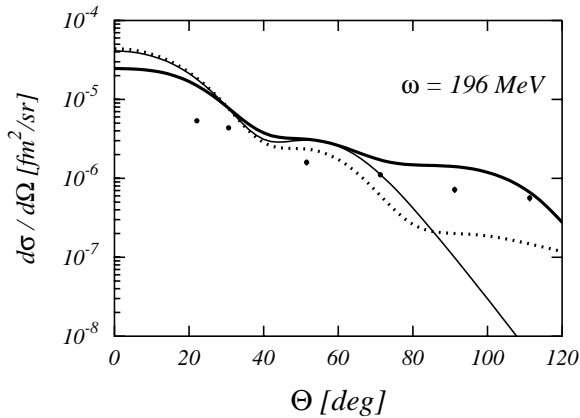}
\caption{Cross sections of the  $^{16}$O($\gamma$,p)$^{15}$N process 
  as a function of the proton emission angle compared
  with the experimental data~\protect\cite{ada88}.
  The thin full line shows the result obtained with OB currents only.
  The inclusion of SRC produces the dotted line. When also the MEC are
  included, the thick full line is obtained. 
}
\label{fig:gpox}
\end{center}
\end{figure}

A relatively large effect produced by the SRC has been found in the
($\gamma$,p) reaction on the $^{16}$O nucleus~\cite{giu03b,ang02}.  In
Fig. \ref{fig:gpox} the cross sections of this process, calculated
with OB currents only (thin full line), with the inclusion of SRC
(dotted line) and by further adding the MEC (thick full line) are
compared with the experimental data~\cite{ada88}.  The relative effect
produced by the SRC on the OB cross section is quite large.
Unfortunately, in this kinematic region, the effects of the MEC are
even larger.

\subsection{Two-nucleon emission: (e,e'NN) and ($\gamma$,NN)}
\label{sec:eepp}
In the processes discussed so far, the effects of the SRC correlations
have been obscured by the presence of the uncorrelated OB terms, or by
the MEC currents.  It is possible to eliminate the contribution of the
OB terms by considering processes where two nucleons are emitted and
detected.  In this case the only mechanism competing with the SRC are
the MEC.  When two-like nucleons are emitted, the only terms of the
MEC contributing to the cross section are the $\Delta$-current
diagrams, the (c) diagrams of Fig. \ref{fig:mec}, where the exchanged
pion is chargeless.

From the theoretical point of view the description of the two-nucleon
emission processes has been treated as a straightforward extension of
the single nucleon emission case.

In the Pavia approach Eq.(\ref{eq:pavia}) is extended as:
\begin{equation}
W({\bf q}) = \int d^3 r\, e^{i{\bf q}\cdot {\bf r}_{12}} \,\,
             \chi_1({\bf r}_{1})\chi_2({\bf r}_{1})J({\bf r}_{12}) \,
   <\Psi^{A-2} ({\bf r}_{12})\mid\Psi^A_0 > 
\end{equation}
with the obvious meaning of the symbols. Now the quantity containing
the correlations is the two body-overlap function, between the target
ground state and the state with A-2 nucleons.

In the Lecce approach the transition density of Eq. (\ref{eq:lecce})
is calculated by using a Slater determinant $\mid\Phi_f > $ with two
particles in the continuum and, obviously, two holes.

In both approaches the interaction between the two emitted nucleons is
not considered. This problem has been investigated by the Pavia
group~\cite{sch03,sch04}, by using an approximation.  The interaction
between the two emitted nucleons has been considered, as has the
interaction between each nucleon and the A-2 nucleus.  The
simultaneous interaction of two nucleons between themselves and with
the remaining A-2 nucleus has been neglected.  This would be a genuine
three-body problem.  The results obtained considering the $^{16}$O
nucleus as a target show that the interaction effect is relevant in
(e,e'pp) reaction, but it is negligible in (e,e'pn) reaction. More
interesting is the fact that in the ($\gamma$,pp) reaction the effect
is always negligible.

\begin{figure}[ht]
\begin{center}
\includegraphics[scale=0.8]{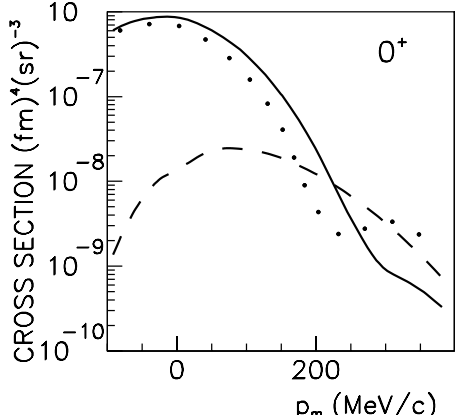}
\caption{Cross sections of the  $^{16}$O(e,e'pp)$^{14}$C process 
  leading to the ground state of the $^{14}$C,
  as a function of the initial momentum of the emitted pair.
  The dadhed line has been calculated by considering the
  $\Delta$-current only. The dotted line with the SRC only, and the
  full line with both contributions.
}
\label{fig:giustiee2p}
\end{center}
\end{figure}

In order to obtain information on the SRC it is necessary to
disentangle the two nucleon emission induced by the correlations from
that produced by the MEC.  We have already seen that the emission of
two-like nucleons eliminate the MEC seagull and pionic diagrams, and
also part of the $\Delta$-current terms.  It is possible to find
kinematic situations where the SRC dominate on the remaining
$\Delta$-current terms~\cite{bar04a}, as is shown in Fig.
\ref{fig:giustiee2p}.  When the $^{16}$O(e,e'pp)$^{14}$C reaction
leads to the ground state of the $^{14}$C nucleus, the
$\Delta$-currents contribution is much smaller than that of the SRC.
The situation is reversed when the nuclear final state is the excited
1$^+$ state in $^{14}$C.

A detailed study of the momentum dependence of the $\Delta$-current
contributions~\cite{ang04} shows that they are minimized at small
values of the momentum transfer. In photon reactions these
contributions are much smaller than those of the SRC for all the
$^{14}$C final states. From this point of view the two-proton emission
induced by real photons with energy far from the $\Delta$ resonance
peak, is the ideal tool to investigate SRC~\cite{ang04}.

The role of the tensor terms of the SRC is however quenched in the
emission of two-like nucleons. The contribution of these terms is
significant only in (e,e'pn) processes~\cite{bar04a}.

\section{Neutrino-nucleus interaction}
\label{sec:nu}
\begin{figure}[ht]
\begin{center}
\includegraphics[scale=0.65]{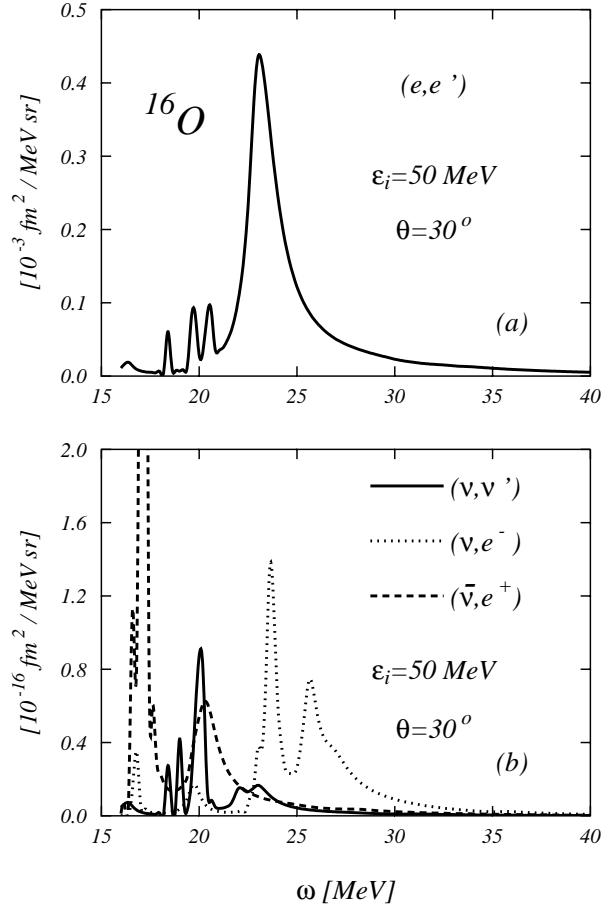}
\caption{Comparison between electron (above) and neutrino (below)
  scattering cross sections. The energy of the incoming lepton and the
  scattering angle are the same in all the reactions considered.
}
\label{fig:nunu}
\end{center}
\end{figure}
In Fig. \ref{fig:nunu} the electron scattering cross section is 
compared with that of neutrino scattering in the same kinematic
conditions. All the cross sections have been calculated for the same 
energy of the projectile and the same scattering angle. The nuclear
transitions have been calculated by using the continuum RPA.
 
The three cross sections have quite a different behavior.  This is
expected for the charge exchange reactions but it is surprising when
electron scattering and neutrino charge conserving neutral current
reactions are compared. In this case the particle-hole configuration
space describing the nuclear excitation is the same in both processes.

The reason for this difference can be traced by making a multipole
decomposition of the cross sections.  In the electron scattering case
the 1$^-$ excitation is responsible for the 98\% of the total cross
section. On the contrary, in the ($\nu$,$\nu'$) case the 1$^-$
contributes only to the 33\% of the cross section, while the main
contribution, 58\%, is due to the the 2$^-$ multipole~\cite{bot04a}.

This result is due to the fact that in the neutrino cross section the
main contributions are given by the transverse axial vector term of
the current operator. This operator excites both natural and unnatural
parity states. In electron scattering the main contribution is due to
the charge operator exciting natural parity states only.  The
dominance of the axial vector term is a quite general result.  Also
the charge exchange reactions are dominated by this term of the
current, and this is the dominant term for all the neutrino cross
sections also in the quasi-elastic region~\cite{bot04b}.

As a consequence of this, it is necessary to be careful in relying on
the fact that a good description of electron scattering data implies a
good description of the neutrino-nucleus cross section.  In spite of
this warning, electron scattering is still the best guide we have to
determine the prediction power of our nuclear models.  The extension
to the neutrino scattering of the electron scattering formalism is
quite straightforward. In these last two years, almost all the Italian
groups working in electron scattering have applied their techniques to
calculate neutrino-nucleus cross-sections.  The calculations have been
mainly done in the quasi-elastic
region~\cite{alb02,mai03,meu04a,meu04b,ben04a,ben04b,ben03a,co02}.

As I have previously pointed out the main correction to the naif MF
model is due to the FSI.  The FSI calculated with a folding model
model~\cite{ben04b,co02,ble01}, produces a reduction of the total
neutrino-nucleus cross section by about 10\%. This is due to the
fact that the FSI spreads the strength of the response and part of it
is moved to excitation energies kinematically prohibited.

The role of the correlations have been estimated to be relatively
large~\cite{ben04a,ben04b}. The peak of the
$^{16}$O($\nu$,e$^-$)$^{16}$F cross section for 1 GeV neutrinos is
reduced by about 20\%.  This effect has been estimated by comparing
FG results with those obtained by using a correlated spectral function
in a local density approximation (LDA). It will be interesting to
disentangle the LDA effects from those produced by the correlations.

The neutrino-nucleus scattering in the quasi-elastic regime has been
found to be the ideal tool for studying 
the strangeness content of the
nucleon~\cite{alb02}. In this case one has to separate the cross
sections where a neutron is emitted from those where a proton is
emitted.  The calculations done in a FG model show about a 25\%
difference between the results obtained with and without strangeness
in the nucleon form factor~\cite{bar96}.  This large effect has been
recently confirmed by a finite nucleus calculation with a relativistic
MF model~\cite{meu04b,meu04c}.

\section{Conclusions}
\label{sec:conc}
In these last two years, the activity of the Italian community
regarding nuclear physics with electroweak probes has 
concentrated on
the region of the quasi-elastic peak.  In this region,
the nuclear excitation is dominated by the single-particle dynamics,
therefore MF models with OB currents are the starting point of the
description of the nuclear excitation.  Other effects induced by MEC,
FSI, SRC and collective modes, are relatively small, and can be
treated as perturbations of the MF results.  There is a convergence of
the results obtained with different models and techniques, and many of
the uncertainties and problems presented and discussed in the
past~\cite{pac01,cen03,cen01} have been resolved.

The comparison with the experimental data is still problematic.
Concerning the inclusive experiments, the Frascati $^{16}$O inclusive
cross sections are well reproduced~\cite{meu03,co02}, as well as the
$^{40}$Ca MIT-Bates separated responses~\cite{meu03,ama94}.  The same
models are unable to describe the the Saclay $^{40}$Ca and $^{12}$C
separated responses.  I think that at present the major problems are
on the experimental side. Experiments to obtain separated responses
in $^{16}$O are necessary, and the incompatibility between the
$^{40}$Ca data measured at MIT and at Saclay should be clarified.

The situation regarding one-nucleon emission processes is slightly
clearer. The same MF models used to describe inclusive experiments are
able to reproduce extremely well the shapes of the cross
sections~\cite{iva02}. The open problem is the understanding of the
spectroscopic factor needed to obtain quantitative agreement with the
data.  Contrary to some claims~\cite{lap00,fra01} it has been shown 
that the spectroscopic factor is a number~\cite{ama03a,rad02}, model
dependent, but independent form momentum and energy transfer.

The search for effects induced by SRC has shown that the best way of
studying them is the use of two-nucleon emission processes. Since the
$\Delta$-current terms become small at low momentum transfer, in the
two-nucleon emission processes real photons seem to be a better tool
than electrons~\cite{giu03b,ang04}. In this field the various
approaches provide similar results, indicating that the theoretical
uncertainties are kept under control.

Even though the neutrino scattering is dominated by the axial part of
the current, absent in electron scattering, the above mentioned
results suggest that the same models could provide a good
description of the quasi-elastic neutrino scattering. This is a  
very important information for the existing and planned experiments
which use the nucleus as a detector to investigate the properties of
the neutrinos and their sources. 

Furthermore neutrino-nucleus experiments in the quasi-elastic region,
are perhaps the best tool we have for information on the strange
content of the nucleon~\cite{alb02,meu04b}.  The conservative estimate
of the nuclear structure uncertainties, that resulted to be of the
same order of the searched effects, should be updated. In our present
understanding of the quasi-elastic excitation this uncertainty is
reduced, therefore the strangeness effects could be identified.

Electron scattering off nuclei is a precision tool to control and
investigate our understanding of nuclear structure. The evaluation of
these cross sections should be used as a benchmark to verify the
validity of approaches aimed at predictions in other nuclear physics
fields. In these last few years the theories have improved remarkably,
and there are signs indicating the possibilities of even greater
improvements in the coming years.

It is however disappointing to notice that many of the experimental
data are old, incomplete, and quite often not accurate enough to
disentangle interesting effects. I join G. Orlandini~\cite{orl04} by
pointing out the need of a major experimental program to investigate
medium-heavy nuclei with electromagnetic probes.
 
%%%%%%%%%%%%%%%%%%%%%%%%%%%%%%%%%%%%%%%%%%%%%%%%%%%%%%%%%%%%%%%%%%%%%%%%%%


\begin{thebibliography}{0}
%
%TORINO
%
\bibitem{alb02} W.M. Alberico, S.M. Bilenky, C. Maieron
    {\it Phys. Rep.} {\bf 358}, 227 (2002).

\bibitem{alb04a}  W.M. Alberico, S.M. Bilenky,
    {\it Phys. Part. Nuclei} {\bf 35}, 297 (2004).

\bibitem{ama02a} J.E. Amaro, M.B. Barbaro, J.A. Caballero,
                  T.W. Donnelly, A. Molinari,
    {\it Eur. Phys. J.} {\bf  A15}, 421 (2002). 

\bibitem{ama02b} J.E. Amaro, M.B. Barbaro, J.A. Caballero,
                  T.W. Donnelly, A. Molinari,
    {\it Phys. Rep.} {\bf 368}, 317 (2002).

\bibitem{ama02c} J.E. Amaro, M.B. Barbaro, J.A. Caballero,
                  T.W. Donnelly, A. Molinari,
    {\it Nucl. Phys.} {\bf A697}, 388 (2002).

\bibitem{ama03a} J.E. Amaro, M.B. Barbaro, J.A. Caballero, F.K. Tabatabaei,
    {\it Phys. Rev.} {\bf C68}, 014604 (2003).

\bibitem{ama03b} J.E. Amaro, M.B. Barbaro, J.A. Caballero,
                  T.W. Donnelly, A. Molinari,
    {\it Nucl. Phys.} {\bf A723}, 181 (2003).

\bibitem{ama04} J.E. Amaro, M.B. Barbaro, J.A. Caballero,
                  T.W. Donnelly, A. Molinari, I. Sick,
    arXiv:nucl-th/0409078

\bibitem{bar04} M.B. Barbaro, J.A. Caballero, T.W. Donnelly, C. Maieron,
    {\it Phys. Rev.} {\bf C69}, 035502 (2004).

\bibitem{dep03} A. De Pace, M. Nardi, W.M. Alberico, 
                 T.W. Donnelly, A. Molinari,
    {\it Nucl. Phys.} {\bf A726}, 303 (2003).

\bibitem{mai03} C. Maieron, M.C. Mart\'{\i}nez, J.A. Caballero, 
                J.M. Ud\'{\i}as,
    {\it Phys. Rev.} {\bf C68}, 048501 (2003). 
%
%PAVIA
%
\bibitem{bar04a} C. Barbieri, C. Giusti, F.D. Pacati, W.H. Dickhoff, 
     {\it Phys. Rev.} {\bf C70}, 014606 (2004). 

\bibitem{gai02} M.K. Gaidarov, K.A. Pavlova, A.N. Antonov, C. Giusti, 
                S.E. Massen, C.C. Moustakidis, K. Spasova,
   {\it Phys. Rev.} {\bf C66}, 064308 (2002).

\bibitem{giu02} C. Giusti, F.D. Pacati,
    {\it Nucl. Phys.} {\bf A699}, 57c (2002).

\bibitem{giu03a} C. Giusti,
    {\it Eur. Phys. J.} {\bf  A17}, 419 (2003). 

\bibitem{giu03b} C. Giusti, F.D. Pacati,
   {\it Phys. Rev.} {\bf C67},  044601(2003).

\bibitem{iva02} M.V. Ivanov, M.K. Gaidarov, A.N. Antonov, C. Giusti,
    {\it Nucl. Phys.} {\bf A699}, 336c (2002).

\bibitem{meu02a} A. Meucci, C. Giusti, F.D. Pacati,
   {\it Phys. Rev.} {\bf C66}, 034610 (2002).

\bibitem{meu02b} A. Meucci, 
   {\it Phys. Rev.} {\bf C65},  044601 (2002).

\bibitem{meu03} A. Meucci, F. Capuzzi, C. Giusti, F.D. Pacati,
   {\it Phys. Rev.} {\bf C67}, 054601 (2003). 

\bibitem{meu04a} A. Meucci, C. Giusti, F.D. Pacati,
    {\it Nucl. Phys.} {\bf A739}, 277 (2004).

\bibitem{meu04b} A. Meucci, C. Giusti, F.D. Pacati,
    arXiv:nucl-th/0405004.

\bibitem{rad03} M. Radici, A. Meucci, W.H. Dickhoff,
   {\it Eur. Phys. J.} {\bf  A17}, 65 (2003). 

\bibitem{rad02} M. Radici, W.H. Dickhoff, E.R. Stoddard,
   {\it Phys. Rev.} {\bf C66}, 014613 (2002).

\bibitem{sch03} M. Schwamb, S. Boffi, C. Giusti, F.D. Pacati,
   {\it Eur. Phys. J.} {\bf  A17}, 7 (2003). 

\bibitem{sch04} M. Schwamb, S. Boffi, C. Giusti, F.D. Pacati,
    {\it Eur. Phys. J.} {\bf  A20}, 233 (2004). 
%
%ROMA 
%
\bibitem{ben04a} O. Benhar,
    arXiv:nucl-th/0408045. 

\bibitem{ben04b} O. Benhar, N. Farina, 
    arXiv:nucl-th/0407106. 

\bibitem{ben03a} O. Benhar,
     arXiv:nucl-th/0307061.
%
% TRENTO
%
\bibitem{bac02} S. Bacca, M.A. Marchisio, N.  Barnea, W. Leidemann, 
                G. Orlandini,  
    {\it Phys. Rev. Lett.} {\bf 89} 052502 (2002).

\bibitem{bac04b} S. Bacca, H. Arenhoevel, N.  Barnea, W. Leidemann, 
               G. Orlandini,
      arXiv:nucl-th/0406080.
 
\bibitem{bac04a} S. Bacca, N.  Barnea, W. Leidemann, G. Orlandini,  
      {\it Phys. Rev.} {\bf C69}, 057001 (2004).

\bibitem{orl04} G. Orlandini,
    {\it Nucl. Phys.} {\bf A737}, 210 (2004).
%
%LECCE
%
\bibitem{ang02} M. Anguiano, G. Co', A.M. Lallena, S.R. Mokhtar,
    {\it Ann. Phys. (NY)} {\bf 296}, 235 (2002).

\bibitem{ang03} M. Anguiano, G. Co', A.M. Lallena,
    {\it J. Phys. G: Nucl. Part.}  {\bf 29}, 1119 (2003).

\bibitem{ang04} M. Anguiano, G. Co', A.M. Lallena,
    arXiv:nucl-th/040841.

\bibitem{bot04a} A. Botrugno, G. Co'
    arXiv:nucl-th/0409041.

\bibitem{co02} G. Co', C. Bleve, I. De Mitri, D. Martello,
    {\it Nucl. Phys. B: Proc. Suppl.} {\bf 112}, 210 (2002).

\bibitem{lal04} A.M. Lallena, M. Anguiano, G. Co',
    arXiv:nucl-th/0407112.

%\bibitem{alb04b} W.M. Alberico,
%    {\it Acta Phys. Pol. B} {\bf 35}, 929 (2004).
%    Alvarez-Ruso L, Barbaro MB, Donnelly TW, Molinari A
%    Nuclear response functions for the N-N*(1440) transition 
%    NUCL PHYS A 724 (1-2): 157-177 AUG 25 2003

%\bibitem{ben04c} O. Benhar, S. Fantoni, G.I. Lykasov, 
%                 U. Sukhatme, V.V. Uzhinsky,
%   {\it Eur. Phys. J.}{\bf  A19}, 147 (2004). 
%
%
%    Benhar O
%    Scaling in many-body systems and proton structure function
%    INT J MOD PHYS B 17 (28): 5139-5149 Part 1 NOV 10 2003
%
%    Benhar O, Fantoni S, Lykasov GI, Sukhatme U
%    Q(2)-dependence of backward pion multiplicity in 
%    neutrino-nucleus interactions
%    PHYS LETT B 527 (1-2): 73-79 FEB 14 2002
%%%%%%%%%%%%%%%%%%%%%%%%%%%%%%%%%%%%%%%%%%%%%%%%%%%%%%%%%%%%%%%%%%%%%%%

\bibitem{bof96} S. Boffi, C. Giusti, F.D. Pacati, M. Radici,
              {\it Electromagnetic Response of Atomic Nuclei},
              Clarendon, Oxford, 1996.

\bibitem{wal75} J.D. Walecka, in {\it Muon Physica},
              Academic Press, New York, 1975. 

%%%%%%%%%%%%%%%%%%%%%%%%%%%%%%%%%%%%%%%%%%%%%%%%%%%%%%%%%%%%%%%%%%%%

\bibitem{leu94} M. Leuschner {\it et al.}, 
              {\it Phys. Rev.} {\bf C49}, 955 (1994).

\bibitem{meu01} A. Meucci, C. Giusti, F.D. Pacati,
              {\it Phys. Rev.} {\bf C64}, 014604 (2001).

%%%%%%%%%%%%%%%%%%%%%%%%%%%%%%%%%%%%%%%%%%%%%%%%%%%%%%%%%%%%%%%%%%%%

\bibitem{ang02a} M. Anguiano and G. Co',
              {\it Jour. Phys. G} {\bf 27},2109 (2001).

\bibitem{hyd87} C.E. Hyde-Wright {\it et al.}, 
              {\it Phys. Rev.} {\bf C35},  880 (1987).

\bibitem{mok00} S.R. Mokhtar, G. Co', A.M. Lallena,
              {\it Phys. Rev.} {\bf C62}, 067304 (2000).

\bibitem{ahr77} J. Ahrens {\it et al.},
              {\it Nucl. Phys.} {\bf A251}, 479 (1975).
%
\bibitem{kie04} A. Kievsky, these proceedings.

\bibitem{qua04} S. Quaglioni {\it et al.}, these proceedings.
%
\bibitem{ang96} M. Anghinolfi {\it et al.}, 
              {\it Nucl. Phys.} {\bf A602}, 405 (1996).

\bibitem{meu04c} A. Meucci, F. Capuzzi, C. Giusti, F.D. Pacati,
              these proceedings

\bibitem{wil97} C.F. Williamson {\sl et al.}, 
              {\it Phys. Rev.} {\bf C56},  3152 (1997).

\bibitem{mez84} Z. Meziani {\sl et al.}, 
              {\it Phys. Rev. Lett.} {\bf 52}, 2130(1984).

\bibitem{mez85} Z. Meziani {\sl et al.}, 
              {\it Phys. Rev. Lett.} {\bf 54}, 1233 (1985).

\bibitem{bar83} P. Barreau {\sl et al.}, 
              {\it Nucl. Phys.} {\bf A402}, 515 (1983).

\bibitem{cap91} F. Capuzzi, C. Giusti, F.D. Pacati, 
              {\it Nucl. Phys.} {\bf A524}, 681 (1991).

\bibitem{ama94} J.E. Amaro, G. Co', A.M. Lallena, 
              {\it Nucl. Phys.} {\bf A578}, 365 (1994).

\bibitem{ama93} J.E. Amaro, G. Co', A.M. Lallena, 
              {\it Ann. Phys. (NY)} {\bf 221}, 306 (1993).

%%%%%%%%%%%%%%%%%%%%%%%%%%%%%%%%%%%%%%%%%%%%%%%%%%%%%%%%%%%%%%%%

\bibitem{mok01} S.R. Mokhtar, M. Anguiano, G. Co', A.M. Lallena, 
              {\it Ann. Phys. (NY)} {\bf 292}, 67 (2001).

\bibitem{iva01} M.V. Ivanov, M.K. Gaidarov, A.N. Antonov, C. Giusti,
    {\it Phys. Rev.} {\bf C64},  014605 (2001).

\bibitem{ada88} G.S. Adams {\it et al.},
    {\it Phys. Rev.} {\bf C38}, 2771 (1988).

\bibitem{bot04b} A. Botrugno,
    {\it Ph. D. Thesis}, Universit\`a di Lecce, unpublished.

\bibitem{ble01} C. Bleve {\it et al.}
    {\it Astr. Phys.} {\bf 16} 145 (2001).

\bibitem{bar96} M. Barbaro, A. De Pace, T.W. Donnelly, 
                A. Molinari, M.J. Musolf, 
    {\it Phys. Rev.} {\bf C54}, 1954 (1996). 

%%%%%%%%%%%%%%%%%%%%%%%%%%%%%%%%%%%%%%%%%%%%%%%%%%%%%%%%%%%%5

\bibitem{pac01} F.D. Pacati,
    {\it Proc. of the 8th Conference on Problems in Theoretical
    Nuclear Physics}, Cortona (Italy), 
    World Scientific, (Singapore) 2001.

\bibitem{cen03} R. Cenni
    {\it Proc. of the 9th Conference on Problems in Theoretical
    Nuclear Physics}, Cortona (Italy), 
    World Scientific, (Singapore) 2003.

\bibitem{cen01} R. Cenni (Ed.),
    {\it Electromagnetic Response Functions of Nuclei},
    Nova Science, Huntington (New York), (2001).

\bibitem{lap00} L. Lapik\'as, G. van der Steenhoven, L. Frankfurt,
                M. Strikman, M. Zhalov,
    {\it Phys. Rev.} {\bf C61}, 064325 (2000). 

\bibitem{fra01} L. Frankfurt, M. Strikman, M. Zhalov,             
    {\it Phys. Lett.} {\bf B503}, 71 (2001). 
%
%
\end{thebibliography}
\end{document}